# MedViz: An Agent-based, Visual-guided Research Assistant for Navigating Biomedical Literature

Huan He[1], Xueqing Peng[1], Yutong Xie[2], Qijia Liu[2], Chia-Hsuan Chang[1], Lingfei Qian[1], Brian Ondov[1], Qiaozhu Mei[2], Hua Xu[1,+]

[1]Department of Biomedical Informatics & Data Science, School of Medicine, Yale University
[2]School of Information & College of Engineering, University of Michigan

+ Correspondence to hua.xu@yale.edu

## Introduction

In the rapidly evolving field of biomedicine, researchers face the challenge of navigating millions of publications spanning diverse knowledge domains[1]. Traditional literature search engines primarily present results as ranked text lists, offering limited support for global exploration, contextual understanding, or sensemaking across a field. Although users can browse individual articles, these list-based interfaces provide little assistance for understanding the overall structure of a research area, identifying thematic relationships, or examining how topics evolve over time, making it difficult to address complex or exploratory research questions.

Recent advances in generative artificial intelligence (AI) and large language models (LLMs), such as ChatGPT[2] and DeepSeek[3], have demonstrated impressive capabilities in information extraction[4], summarization[5], reasoning, and question answering[6]. However, most existing LLM-based or agent-based tools operate through dialog-centric interfaces that return text-only responses generated with a limited number of publications. These chatbot-style systems are typically detached from literature exploration workflows and provide limited support for navigating the complete document collections, discovering global patterns, or iteratively refining analytical context through user interaction[7,8]. As a result, researchers are often confined to small, agent-selected subsets of literature, without visibility into the broader knowledge landscape from which these results are drawn.

To address these limitations, we developed **MedViz**, a visual analytics system that integrates interactive visualization with context-aware, multi-agent AI to support large-scale literature exploration. MedViz transforms literature search from list-based retrieval into space-based semantic sensemaking by visualizing the entire semantic space of a literature corpus. This approach allows researchers to perceive the global structure of a field, identify topic clusters, recognize underexplored or densely studied areas, and explicitly construct analytical context through visual selection. Question answering, summarization, and analysis in MedViz are therefore grounded in a transparent, user-defined subset of the literature, rather than opaque agent-driven retrieval alone.

## System Overview

As illustrated in Figure 1, MedViz consists of three tightly integrated components: a scalable data processing pipeline (Fig. 1a), an interactive visual analytics interface (Fig. 1b–c), and a context-aware, agent-based reasoning framework (Fig. 1d).

***Data processing pipeline***. Users begin by providing a user-specific list of articles, such as a PubMed query result or curated collection. As shown in Figure 1a, MedViz first aggregates bibliographic metadata, abstracts, and citation information from external sources including PubMed and related literature services (e.g., CrossRef and iCite). Article texts are embedded into high-dimensional semantic vectors using pretrained text embedding models. These embeddings are projected into two-dimensional coordinates via dimensionality reduction, enabling spatial visualization. Hierarchical clustering is then applied to identify coherent topical groupings, and large language models are used to automatically generate human-readable topic labels. Finally, citation and co-citation relationships are processed and visually simplified through network edge bundling, producing a semantic map that supports both topical and relational exploration.

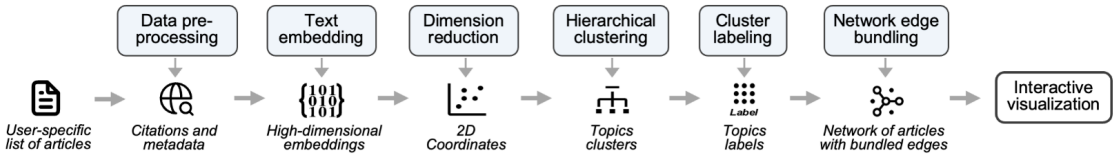
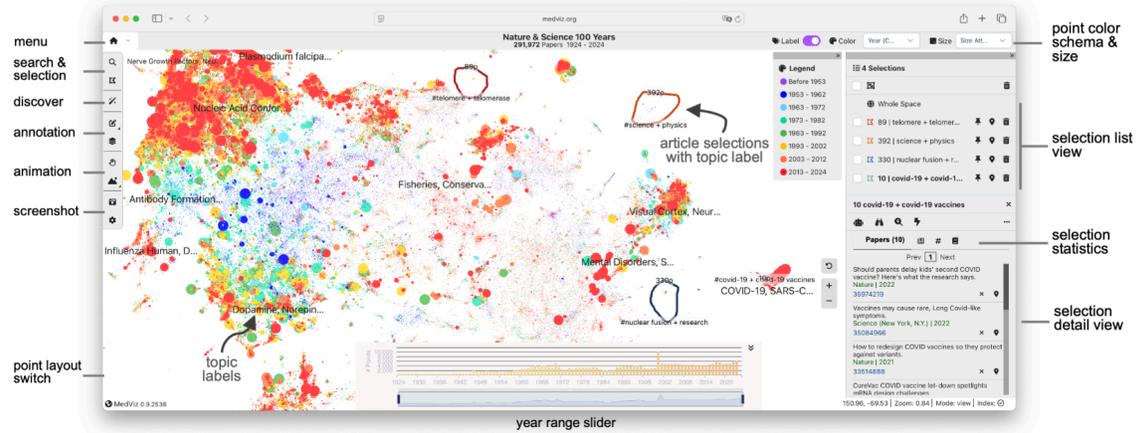
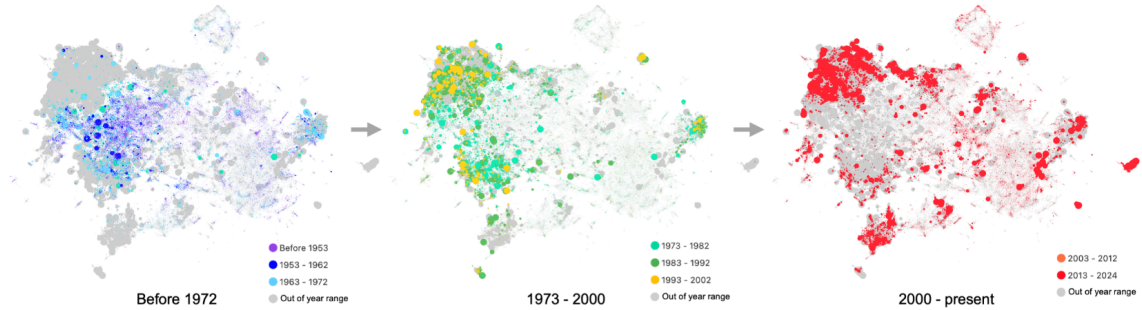
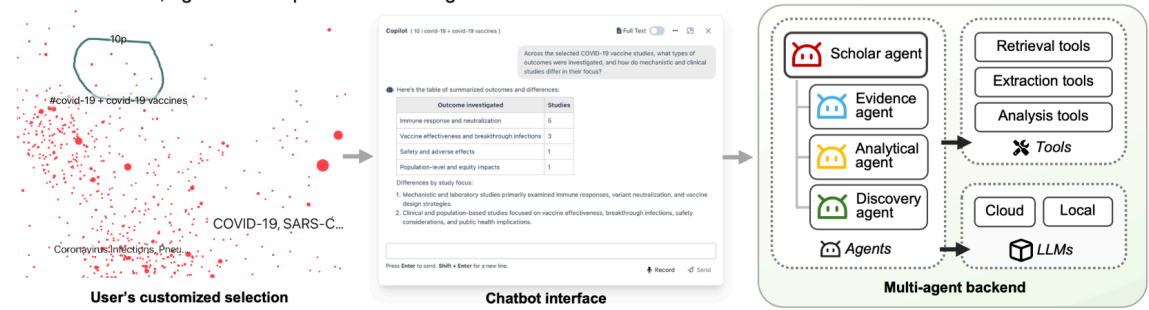

**Figure 1**. Overview of the MedViz system for space-based exploration of biomedical literature. a) MedViz pipeline embeds articles into a high-dimensional semantic space, projects them to 2D, clusters them topically, and labels them with LLMs. b) The literature corpus is visualized as an interactive semantic point cloud, where spatial proximity reflects semantic similarity. Users can search, filter, and visually select regions or articles to define explicit analytical context. c) Temporal dynamics are explored by filtering the same semantic space by publication year, enabling comparison of topic evolution without recomputing the layout. d) Visual selections provide context for agent-based question answering. A Scholar agent orchestrates specialized agents for evidence extraction, analysis, and discovery, integrating AI-assisted reasoning with interactive visualization.

*Visual analytics interface*. The resulting semantic map is rendered in the MedViz web-based interface as a large-scale point cloud, where each point represents an individual article and spatial proximity reflects semantic similarity (Figure 1b). Points can be visually encoded by publication year, citation count, or other metadata attributes, while bundled edges reveal co-citation structure across subfields. Built using WebGL and Three.js[9], the interface supports interactive filtering, keyword-based highlighting, region selection, and cluster-level inspection up to one million of points. Unlike traditional search interfaces, MedViz enables researchers to interpret search results within a global reference frame. Users can visually identify regions or clusters of interest, delineate subsets of articles directly on the semantic map, and inspect topic labels that summarize the dominant themes of each cluster. These visual selections define an explicit analytical context that can be refined iteratively, supporting deeper exploration and comparison across related areas. Moreover, MedViz further supports analysis of how research topics evolve over time. As shown in Figure 1c, the same semantic space can be interactively filtered by publication year, allowing users to observe temporal shifts in topic prominence and spatial distribution without recomputing the underlying layout. By visualizing different time slices within a consistent semantic coordinate system, MedViz enables intuitive comparison of historical and contemporary research trends, revealing paradigm shifts, emerging topics, and declining areas of interest.

*Context-aware, agent-based reasoning*. To complement visual exploration, MedViz integrates a context-aware, multi-agent reasoning framework that operates directly on user-defined selections (Figure 1d). Rather than autonomously searching the literature, MedViz agents are grounded in the subset of articles explicitly selected by the user through visual or keyword-based interaction. At the core of this framework is a Scholar agent, which serves as an orchestrator and planner. The Scholar agent interprets user questions in conjunction with the current visual context (e.g., selected articles, regions, clusters, and time ranges), decomposes complex requests into subtasks, and coordinates specialized agents and tools to execute them. These specialized agents include an Evidence agent for retrieval and structured extraction of information from selected papers, an Analytical agent for aggregation, comparison, and statistical analysis (e.g., trend analysis or meta-analytic summaries), and a Discovery agent for identifying patterns, gaps, or hypothesis-generating insights within the selected semantic region.

Importantly, agents in MedViz are not limited to returning text responses. Through tool interfaces, agents can also issue executable commands to the visual analytics interface, such as highlighting relevant clusters, annotating topic labels, adjusting temporal filters, or pinning representative articles for inspection. This tight coupling between reasoning and visualization enables an iterative sensemaking loop in which users can refine selections, pose follow-up questions, and immediately see the effects of agent-assisted analysis in the semantic map.

## Applications

MedViz supports a range of exploratory and analytical workflows that extend beyond conventional literature search. By providing a global view of a research domain, MedViz allows investigators to rapidly familiarize themselves with unfamiliar or fast-evolving fields, identify major thematic clusters, and understand how research trajectories shift over time. For example, a researcher entering the field of cancer immunotherapy can quickly locate clusters related to checkpoint inhibition, tumor microenvironment, or vaccine-based approaches, and then focus analysis on specific subsets of interest.

Beyond landscape overview, MedViz enables targeted, selection-driven analysis. Users can delineate clusters to generate focused summaries, compare thematic regions across time, or extract structured evidence from selected papers to support rapid or mini systematic reviews. The ability to visually contextualize evidence also aids in identifying underexplored connections, avoiding duplication of effort, and assessing the novelty of new research ideas. By grounding downstream analysis in an explicitly defined semantic context, MedViz supports more transparent and controllable reasoning than agent-driven search alone.

**Conclusion and future work**

MedViz introduces a new paradigm for biomedical literature exploration by integrating large-scale visual analytics with context-aware, multi-agent AI. By visualizing the entire semantic space of a literature corpus, MedViz enables researchers to move beyond static search results and engage in space-based sensemaking, constructing analytical context through direct interaction with the knowledge landscape. Agent-based reasoning in MedViz is explicitly grounded in user-selected subsets, supporting evidence-backed summarization, extraction, and analysis while preserving transparency and user control.

MedViz is designed to be flexible and extensible, supporting both public datasets and user-provided collections, with deployment options ranging from hosted services to containerized local instances. Future work will explore tighter integration with knowledge graphs, support for multimodal biomedical data, and expanded analytical capabilities for evidence synthesis. MedViz is freely available at https://medviz.org, providing the biomedical research community with a browser-based environment for interactive knowledge discovery and scientific insight.

# Supplementary Note

## Motivation

A fundamental challenge in biomedical research is the sheer scale and complexity of the scientific literature. [1,2] For many established or rapidly evolving research areas, the number of relevant publications can easily reach tens or hundreds of thousands. While individual studies may address specific aspects of a research question, synthesizing evidence, identifying overarching trends, and forming a coherent understanding across such large corpora remain difficult and time-consuming. [3] As a result, answering even moderately complex research questions often requires substantial manual effort to locate, filter, and read relevant studies.

***First, the scale of modern scientific literature exceeds the limits of sequential, list-based exploration***. Traditional search workflows assume that users can iteratively review ranked lists of articles and gradually build understanding through reading. [4] This assumption breaks down at scale, where the volume and diversity of publications make it difficult to gain an overview of a field or to systematically integrate evidence across studies.

***Second, list-based literature search systems provide limited support for understanding the global structure of a research domain***. Although literature systems, such as Semantic Scholar, [5] excel at retrieving articles ranked by keyword relevance or recency, they present results as isolated items rather than as part of a coherent knowledge landscape. Users receive little insight into how topics relate to one another, which subareas are densely studied or underexplored, or how research themes shift over time. Even when basic temporal statistics are available, they do not reveal how scientific focus evolves conceptually or spatially across a field.

***Third, recent LLM-powered and agent-based literature tools introduce new opacity into the exploration process***. Recent literature review systems, such as Elicit [6] and Scite [7] autonomously retrieve and process a small subset of articles, generating concise textual responses without exposing the underlying selection process. This obscures potential retrieval bias, limits users' visibility into the broader literature landscape, and makes it difficult to assess coverage, explore alternative contexts, or verify whether important subtopics have been overlooked. As a result, users are often presented with answers derived from narrow and opaque samples, with limited support for iterative refinement or context-aware analysis.

## System Architecture

To enable scalable, field-level sensemaking, MedViz adopts a semantic space representation that provides an interpretable global overview while preserving meaningful local neighborhoods. To make exploration transparent and controllable, MedViz treats user interaction as the primary mechanism for constructing analytical context: users define inspectable subsets of the corpus through map-based selection, cluster inspection, and temporal filtering. Finally, to support deeper reasoning without introducing hidden retrieval bias, MedViz integrates AI agents that operate exclusively on user-defined selections, enabling evidence extraction, synthesis, and analysis grounded in explicit visual context rather than opaque autonomous retrieval.

MedViz adopts a modular, layered system architecture that integrates large-scale data processing, agent-based reasoning, and interactive visual analytics into a unified workflow (Supplementary Figure S1). The system is organized into three primary layers:

**1) a data processing pipeline** for constructing semantic representations of user-specific literature collections. This pipeline transforms a user-specified list of articles (e.g., the result of a PubMed query) into a structured dataset suitable for visualization and downstream analysis. Starting from bibliographic metadata and citation information, the pipeline performs data pre-processing, text embedding, dimensionality reduction, hierarchical clustering, topic label generation, and citation network edge bundling. The output dataset serves as a shared foundation for both visual analytics and agent-based reasoning.

2) **a backend service** layer that coordinates agent-based analysis and external tools. At the center of the backend is an agent-based architecture designed to support flexible, task-oriented analysis. These agents can be invoked individually or composed dynamically in response to user requests.

3) **a web-based visual analytics interface** that supports large-scale visualization and interactive exploration. By tightly coupling visual exploration with agent-driven analysis, MedViz maintains contextual continuity between what users see, select, and query throughout the exploration process.

Technical details for each layer will be described in the following sections.

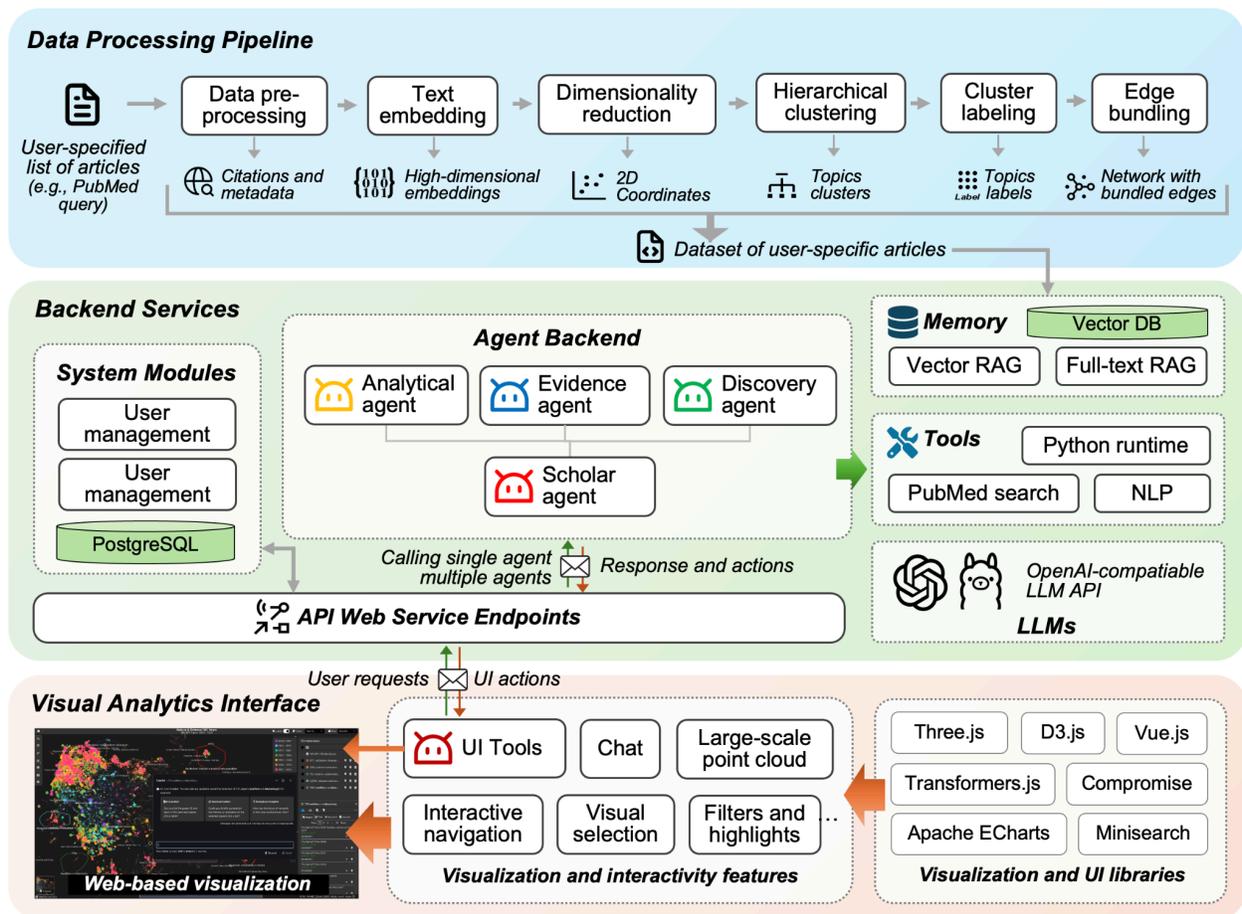

**Supplementary Figure S1**: Overall system architecture. The offline pipeline transforms an input corpus into map-ready artifacts (2D coordinates, metadata, clusters, and topic labels) and optional analysis-ready artifacts (embeddings/indexes and network relations). The WebGL-based frontend provides interactive exploration and emits structured context payloads representing user-defined selections and filters. Context-aware agent services operate on demand over this explicit context and return both textual analyses and declarative UI action commands, enabling a closed-loop workflow that couples reasoning and visualization.

## Software dependencies

The core software dependencies of MedViz are summarized in the table below, listing key libraries used for each stage of the data pipeline, visualization frontend, and agent services.

| Module | Package / Framework | Description |
| --- | --- | --- |
| Data pipeline | pandas | Tabular data processing and numerical computation. https://pandas.pydata.org/ |
| | NumPy [8] | Python numerical and scientific computing. https://numpy.org/ |
| | sentence_transformers | Generating semantic embeddings for scientific documents https://www.sbert.net/ |
| | LargeVis [9] | Large-scale dimensionality reduction for embedding visualization https://github.com/lferry007/LargeVis |
| | opentsne | Alternative dimensionality reduction for semantic layouts https://opentsne.readthedocs.io/en/stable/ |
| | datashader | Scalable rasterization and aggregation for large point clouds during preprocessing https://datashader.org/ |
| Visualization | WebGL | GPU-accelerated rendering of large-scale semantic point clouds https://www.khronos.org/webgl/ |
| | WebWorkers | Web API for offloading computation to maintain interactive performance https://www.w3.org/TR/workers/ |
| | three.js | High-performance 3D rendering and scene management https://threejs.org/ |
| | d3.js [10] | Visualization utilities for creating https://d3js.org/ |
| | ECharts [11] | Interactive statistical and temporal charts https://echarts.apache.org/ |
| | Vue.js | Component-based frontend framework for UI development https://vuejs.org/ |
| Backend service | FastAPI | Backend API for data access, agent invocation, and system coordination https://fastapi.tiangolo.com/ |
| | OpenAI APIs | LLM-based services for agent reasoning and topic labeling https://openai.com/api/ |

**Supplementary Table S1.** *Core software libraries and web technologies used by MedViz.*

## Data Pipeline Specification

As shown in Supplementary Figure S1, the data pipeline is executed offline to transform an input corpus into a map-ready dataset and supporting indexes. Given a user-provided article list (e.g., PMID list, PubMed query results or a curated bibliography), the pipeline aggregates bibliographic metadata and text (titles/abstracts) and constructs multiple derived products: (i) semantic embeddings for each document, (ii) 2D coordinates via dimensionality reduction for map layout, (iii) cluster assignments (optionally hierarchical) for topic grouping, (iv) LLM-generated topic labels for cluster summarization, and (v) optional citation / co-citation relations for network overlays (e.g., bundled edges).

The pipeline produces a set of artifacts that can be consumed by the frontend and agent services. At minimum, MedViz requires a .tsv file, which contains document identifiers and 2D coordinates, along with metadata used for encoding and filtering (e.g., publication year, citation count). Additional artifacts (e.g., topic labels, embeddings, indexes, and edges, etc.) enable similarity search, context expansion, and agent-assisted retrieval within user-defined subsets.

## Data Sources and Preprocessing

MedViz operates on a user-specific collection of biomedical articles, such as the results of a PubMed query or a curated literature list of identifiers (e.g., PMID). Bibliographic metadata, including title, abstract, author list, journal, publication date, and unique identifiers (PMID/PMCID), were primarily retrieved from PubMed, using a combination of the Entrez Programming Utilities (E-utilities) API [12] for query-driven access and the PubMed FTP service for bulk retrieval of baseline and update files. When available, full-text articles were obtained from PubMed Central in structured XML format to support reliable text extraction. Citation-related metadata, including citation counts and relative citation ratio (RCR), [13] were collected from iCite to enable impact-aware visual encoding and network construction. [14]

All retrieved records were processed through a unified pre-processing pipeline designed to prepare the data for semantic embedding and downstream visual analytics. Metadata from heterogeneous sources were normalized into a consistent schema, with explicit resolution of duplicated records, missing fields, and identifier mismatches across PubMed, PMC, and iCite. Textual content (titles and abstracts, and full text when available) was cleaned by removing markup, normalizing Unicode characters, and consolidating whitespace. The resulting structured corpus served as input for text embedding. A sample dataset generated after data pre-processing is available in our GitHub repository.

## Semantic Embedding and Dimensionality Reduction

To construct the semantic representation underlying the MedViz literature map, we embed the textual content of each article into a high-dimensional vector space using LLM-based text embeddings. [15] We are using the embedding model, google/embeddinggemma-300m, to generate text embeddings for articles. [16]

By default, the embedding input consists of the article title and abstract concatenated into a single text sequence. This choice reflects a deliberate design decision informed by empirical observations: while metadata such as journal name, publication year, or controlled vocabulary terms (e.g., MeSH) provide valuable contextual signals, including them in the embedding input introduces strong external structure that can dominate the latent space. In particular, embeddings that incorporate metadata tend to produce clusters organized by venue or time period rather than by conceptual similarity. In contrast, embeddings derived solely from title and abstract yield more coherent, topic-driven groupings that better reflect the semantic relationships among studies.

The resulting high-dimensional embeddings are used for similarity computation, while dimensionality reduction is applied solely for visualization. Specifically, embeddings are projected into two-dimensional coordinates using LargeVis, [9] a dimensionality reduction algorithm that first constructs an approximate k-nearest neighbor graph and then optimizes a low-dimensional layout via a probabilistic model and asynchronous stochastic gradient descent. LargeVis preserves local neighborhood structure while maintaining a stable global layout, making it well suited for interactive exploration of large-scale literature maps. In this work, LargeVis was configured using default parameters. The final output of this stage consists of a lightweight tabular representation, where each article is associated with a unique identifier, its high-dimensional embedding, and corresponding two-dimensional coordinates, forming the basis for downstream clustering, topic labeling, network construction, and interactive visualization. In addition, tSNE[17] can also be used in case LargeVis is not available.

To support efficient data loading, rendering, interaction, and downstream analytics, MedViz saves the processed literature corpus using a lightweight, tabular data format. By default, the system expects a tab-separated values (TSV) file in which each row corresponds to a single article and each column encodes metadata, textual content, or visualization attributes. This design allows the semantic map to be rendered directly from a single file, while remaining flexible enough to accommodate alternative datasets beyond biomedical literature. The detailed column definitions are shown in the Supplementary Table S2.

| Column name | Description |
| --- | --- |
| pmid | PubMed identifier serving as the unique article ID |
| date | Publication date (year or full date, depending on availability) |
| journal | Journal name |
| title | Article title |
| abstract | Article abstract |
| mesh_terms | Associated MeSH terms (semicolon- or comma-separated) |
| x | X-coordinate of the article in the 2D semantic map |
| y | Y-coordinate of the article in the 2D semantic map |
| citation_count | Citation count derived from external sources (e.g., iCite) |
| size | Visual size attribute (e.g., relative citation ratio from iCite) |
| color | Visual color attribute (e.g., encoding cluster label) |

***Supplementary Table S2***. *Default TSV schema*

## Clustering and Topic Label Generation

To identify topical structure within the semantic map and provide interpretable summaries at multiple levels of granularity, MedViz applies a hierarchical clustering and topic labeling pipeline operating directly on the two-dimensional embedding space. This design enables both global and local thematic patterns to be revealed while supporting interactive exploration across zoom levels.

**Hierarchical density-based clustering**

We first group publications based on spatial proximity in the 2D semantic map using a density-based clustering strategy. Specifically, we employ Hierarchical Density-Based Spatial Clustering of Applications with Noise (HDBSCAN), [18] which does not require pre-specifying the number of clusters and is robust to noise and outliers. Clustering is performed in a bottom-up, hierarchical manner: initial parameter settings (e.g., minimum cluster size and minimum samples) identify fine-grained clusters, and parameters are progressively relaxed to reveal coarser groupings. As higher-level clusters emerge, lower-level clusters are assigned to parent clusters based on membership overlap criteria, resulting in a hierarchical topic structure represented as a tree. Clusters that do not merge at higher levels remain as root nodes, corresponding to distinct high-level topics. This procedure captures both broad thematic regions and more specific subtopics within the same semantic space. Our team also proposed a new method called TopicForest to facilitate multi-scale exploration and visualization of literature corpora. [19]

**Topic representation and label generation**

To generate human-readable topic labels for each cluster, we leverage term-frequency–based representations. For high-level clusters, we adopt a class-based TF–IDF (c-TF–IDF) approach, aggregating terms across documents within each cluster and contrasting them against the rest of the corpus to identify representative topic descriptors. [20] In the pipeline we primarily use MeSH terms associated with each publication, which provide domain-specific coverage and hierarchical organization. While c-TF–IDF effectively distinguishes major topics, it tends to produce less discriminative labels for closely related sibling sub-clusters. To address this limitation, we apply a tree-based TF–IDF strategy that computes inverse document frequency locally among sibling clusters sharing the same parent node, rather than across the entire corpus. This localized comparison enhances sensitivity to subtle topical differences within densely populated regions of the semantic map, yielding more specific and informative labels for fine-grained subtopics.

# Visual Analytics Interface

## Point and Edge Rendering

Conventional techniques such as SVG or Canvas 2D rely on CPU-bound rendering pipelines, where each graphical element is individually drawn and updated by the browser's layout engine. This approach performs adequately for small datasets (typically under 10,000 elements) but exhibits exponential degradation in frame rate and interactivity as dataset size increases, resulting in slow redraws and lag during panning or zooming.

To overcome these limitations, MedViz employs a WebGL-based GPU rendering approach, implemented through the Points system in Three.js, [21] which is a high-performance JavaScript library that provides a simplified, higher-level interface for creating and rendering 3D graphics directly in the browser using WebGL. Based on WebGL, rather than sending thousands of individual drawing commands to the browser, all publication points are transmitted to the GPU in a single batch operation and rendered simultaneously. This minimizes communication overhead between the CPU and GPU and allows the graphics card to process millions of points in parallel. Visual attributes such as color, transparency, and size are encoded in GPU-resident textures, enabling dynamic updates (e.g., highlighting selected clusters or adjusting visual scales) without re-rendering the full scene.

Visualizing citation and co-citation relationships between publications reveals structural patterns that are not visible from point-based layouts alone. However, directly rendering all pairwise links produces extreme visual clutter and computational overhead because the number of edges typically grows superlinearly with the number of nodes. To address this, MedViz integrates a kernel density estimation (KDE)–based edge-bundling algorithm, known as Hammer Bundle, which aggregates geometrically similar edges into smooth bundled curves. [22] This approach preserves global structural cues, such as topic interconnections or citation pathways, while greatly reducing the number of rendered primitives. Compared with force-directed or hierarchical bundling, KDE-based bundling offers superior scalability for dense, unstructured graphs such as literature networks.

The resulting bundles are stored as polyline segments rather than raw straight edges, reducing the visible edge count by more than an order of magnitude while maintaining the perceptual impression of connectivity. This compression enables the network to be visualized interactively within the WebGL environment. Bundles are rendered as curved splines using GPU-accelerated line materials in Three.js, which allows smooth transitions in opacity and thickness according to bundle density or citation weight. Edges can also be selectively highlighted or filtered based on user-defined criteria (e.g., co-citation strength or temporal range).

## Interactivity Design and Implementation

### Real-time point hover with quadtree spatial indexing

MedViz supports continuous mouse/trackpad hover feedback by displaying the nearest article under the cursor. For small datasets, point picking can be implemented via raycasting (e.g., THREE.Raycaster), which tests intersections between a ray projected from the camera and rendered objects. However, raycasting against large point clouds can incur substantial CPU overhead and becomes perceptibly laggy at million-scale rendering, especially under high-frequency pointer events.

To achieve real-time hover at scale, we decouple interaction queries from the rendering geometry and build a 2D quadtree over projected point coordinates during data loading. Each article point is inserted into the quadtree based on its 2D map location; [23] the time complexity of insertion is $O(\log n)$ on average (tree height), yielding $O(n \log n)$ expected build time for n points, and supports efficient local neighborhood queries. During interaction, the cursor position is mapped to the same 2D coordinate system, and we perform a range query within a small screen-space radius (converted to world/map units). The candidate set returned by the quadtree is typically small; we then compute the nearest point among candidates to determine the hover target. This reduces per-hover-event work from "scan all points" to approximately $O(\log n + k)$, where k is the number of candidates in the local neighborhood.

Meanwhile, this 2D quadtree is also used in the freeform polygon selection. Given a user-drawn polygon, MedViz leverages the quadtree as a spatial acceleration structure to retrieve points within the bounding box. This approach supports near real-time selection updates even for million-scale point clouds.

**Multi-cluster hover/highlight via GPU DataTexture**

To support instant visual feedback when hovering or filtering clusters, MedViz implements a GPU-side highlighting mechanism that encodes highlight state in a DataTexture uploaded to the GPU and referenced in the point shader. In the fragment and vertex shader, each point fetches its highlighting status and renders its visual encoding accordingly (e.g., opacity/size emphasis). Because of the small size of texture needs updating (proportional to the number of clusters rather than the number of points), MedViz can highlight multiple clusters simultaneously as well as other visual effects with a single GPU state update, enabling responsive cluster hover, comparison, and multi-selection without re-uploading million-scale geometry.

## Multi-agent Framework

MedViz integrates a multi-agent framework to support task-oriented reasoning and literature analytics (Supplementary Figure S1). The agent backend is exposed through API web service endpoints and operates over the user-specific literature dataset produced by the data processing pipeline. User requests originate from the visual analytics interface (e.g., selected regions, highlighted clusters, or filtered subsets) and are forwarded to the agent backend together with structured context, including a set of PMIDs and relevant metadata. Agents return both textual responses and optional UI actions (e.g., highlight clusters, filter by criteria, or annotate selections), enabling a closed-loop workflow that preserves analysis context across interaction steps.

The agent backend is implemented using the OpenAI Agents framework, [24] which provides structured tool calling, message routing, and multi-step planning. Agents are invoked either as single specialists for focused tasks or as a coordinated team for composite workflows. A lightweight controller component in the API layer determines which agent(s) to call based on the user request, assembles the appropriate context payload (e.g., selected papers, query text, and interaction state), and manages intermediate results.

### Specialized agents

MedViz provides several specialized agents, each optimized for a distinct class of user tasks. While these agents share the same tool and memory interfaces, they differ in their prompting, retrieval policy, and output structure.

**Scholar agent**

The Scholar agent serves as the general-purpose entry point for user queries and is invoked by default. It performs task analysis, determines whether retrieval is needed, and routes subtasks to specialist agents when appropriate. Typical functions include general question answering over the selected literature, scoping and reframing queries, generating structured plans, and producing citation-backed responses via OpenScholar. The Scholar agent also mediates follow-up requests by maintaining continuity with the current UI context (e.g., the same selected papers or semantic region) to avoid re-specifying constraints.

**Evidence agent**

The Evidence agent is optimized for structured extraction and evidence synthesis from a user-specified subset of papers (e.g., a polygon selection or a set of PMIDs). Given selected documents, it can extract predefined fields (e.g., population, intervention, outcomes, endpoints, study design) and assemble results into machine-readable outputs such as tables or JSON records suitable for downstream comparison and visualization. When key information is missing or ambiguous in the selected set, the Evidence agent may call search tools to retrieve additional context (e.g., related trial reports or full-text segments) and then re-run extraction.

**Discovery agent**

The Discovery agent supports exploratory workflows by identifying semantically related literature beyond the current focus. It performs similarity-based retrieval over the embedding space and vector database to surface "nearest-neighbor" papers, adjacent clusters, or underexplored semantic regions connected to the user's selection. The Discovery agent can return a research hypothesis and optionally propose follow-up directions grounded in similarity evidence.

**Analytical agent**

The Analytical agent performs quantitative analyses over the selected corpus and can generate and execute Python code through an isolated runtime when computation is required. Typical tasks include descriptive statistics (e.g., publication trends, citation distributions), cluster comparisons, sensitivity analyses (e.g., filtering by year/journal and recomputing summaries), and transforming extraction outputs into plots or aggregated tables.

## Retrieval and memory services

To support retrieval-augmented generation (RAG),[25] MedViz integrates a memory layer containing (1) a vector index for embedding-based similarity search and (2) a full-text index for keyword retrieval. The system uses OpenScholar as the primary RAG retriever, which enables citation-aware retrieval and grounding for synthesized answers.[26] Retrieval is performed in two modes depending on the task: in-collection retrieval, restricted to the current user dataset or selected subset; and open retrieval, which expands to external sources (e.g., PubMed/PMC) when the selected corpus is insufficient. Retrieved evidence is returned with identifiers and provenance (PMID/PMCID, title, source type) so agent outputs can be linked back to the underlying documents.

## Function tools

All agents interact with external capabilities through a unified tool interface based on Model Context Protocol (MCP),[27] including PubMed/PMC search, the OpenScholar retriever, NLP utilities, and the Python runtime. Tool calls are explicitly logged and structured, enabling the system to separate model reasoning from deterministic computation. In addition, UI actions (e.g., highlight points, label papers, and navigations) are also included as tools for agents to present results as needed.

# Limitations and Design Trade-offs

While MedViz is designed to support interactive exploration of large-scale biomedical literature, several limitations and design trade-offs remain.

**Scalability and browser-side performance.** MedViz currently supports interactive visualization of up to approximately 1.2 million points within a standard web browser. This upper bound reflects practical constraints imposed by browser memory limits, GPU buffer sizes, and event-handling latency during interaction. To maintain responsive navigation and real-time feedback, we adopt a set of performance-oriented design choices, including GPU-accelerated point rendering, compact data representations, and spatial indexing structures for interaction. Nevertheless, rendering and interacting with datasets substantially larger than this scale remains challenging in a purely browser-based environment. Ongoing work explores additional optimizations, such as more aggressive level-of-detail strategies, progressive loading, and alternative data layouts, to further extend the feasible scale while preserving interactivity.

**Reliability and stability of agent-based reasoning.** MedViz integrates LLM-based agents to support retrieval, extraction, and analysis; however, these agents are subject to known limitations of generative AI, including variability in responses and the risk of hallucinated or incomplete outputs. To mitigate these issues, the system emphasizes RAG and constrained tool usage, particularly for extraction and analysis tasks. Nevertheless, agent behavior may vary across model versions and prompting strategies. Future work will further integrate coding-oriented agents and tighter verification loops to improve robustness and reproducibility.

**Data source coverage and licensing constraints.** At present, MedViz primarily operates on literature sourced from PubMed and associated open resources due to both practical availability and licensing considerations. The current implementation does not directly support imports from third-party reference management software (e.g., EndNote, Zotero, or Mendeley) or proprietary export formats. Nevertheless, the underlying processing pipeline is data-agnostic and can be applied to other user-provided structured text collections when articles and metadata are available in supported formats.

**Supplementary Reference**